\documentclass[prd,aps,amsmath,amssymb, preprintnumbers, reprint,longbibliography,superscriptaddress,nofootinbib]{revtex4-1}

\usepackage{bm,graphicx,wasysym}
\usepackage{ulem}
\usepackage{xcolor}
\usepackage{color}

\allowdisplaybreaks

\newcommand\sumint[1]{\int\kern-1.5em\sum\nolimits_{#1}}

\newcommand\del{\partial}

\newcommand{\ud}{\mathrm{d}}

\def\({\left(}
\def\){\right)}
\def\[{\left[}
\def\]{\right]}

\newcommand{\f}[2]{\frac{#1}{#2}}
\def \bal#1\eal  {\begin{align} #1 \end{align}}
\newcommand{\be} {\begin{equation}}
\newcommand{\ee} {\end{equation}}
\newcommand{\bc}{\begin{center}}
\newcommand{\ec}{\end{center}}
\newcommand{\bim} {\begin{itemize}[noitemsep]}

\newcommand{\eim} {\end{itemize}}

\newcommand{\nn} {\nonumber\\}
\newcommand{\eref}[1]{Eq.~(\ref{#1})}

\newcommand{\pd} {\partial}
\newcommand{\mc} {\mathcal}

\newcommand{\ai}{{\alpha}}

\newcommand{\li}{{\lambda}}

\newcommand{\ei}{{\eta}}
\newcommand{\oi}{\omega}

\begin{document}

\hfill {\footnotesize USTC-ICTS/PCFT-22-33}

\title{Q-ball Superradiance}

\author{Paul M.~Saffin}
\email{paul.saffin@nottingham.ac.uk}
\affiliation{School of Physics and Astronomy, University of Nottingham,\\
University Park, Nottingham NG7 2RD, United Kingdom}

\author{Qi-Xin Xie}
\email{xqx2018@mail.ustc.edu.cn}
\affiliation{Interdisciplinary Center for Theoretical Study, University of
Science and Technology of China, Hefei, Anhui 230026, China}

\author{Shuang-Yong Zhou}
\email{zhoushy@ustc.edu.cn}
\affiliation{Interdisciplinary Center for Theoretical Study, University of
Science and Technology of China, Hefei, Anhui 230026, China}
\affiliation{Peng Huanwu Center for Fundamental Theory, Hefei, Anhui 230026, China}

\begin{abstract}

Q-balls are non-topological solitons that coherently rotate in field space. We show that these coherent rotations can induce superradiance for scattering waves, thanks to the fact that the scattering involves two coupled modes. Despite the conservation of the particle number in the scattering, the mismatch between the frequencies of the two modes allows for the enhancement of the energy and angular momentum of incident waves. When the Q-ball spins in real space, additional rotational superradiance is also possible, which can further boost the enhancements. We identify the criteria for the energy and angular momentum superradiance to occur. 

\end{abstract}

\maketitle

Q-balls arise as solitonic solutions \cite{Rosen:1968mfz,Friedberg:1976me,Coleman:1985ki} in a variety of field theories that admit attractive nonlinear interactions~\cite{Lee:1991ax}. The attractive nature of the interactions allows charges to condense into a localized object, which however is not a static field configuration, as the phase of the field evolves in time, so as to evade Derrick’s theorem~\cite{Derrick:1964ww}. Q-balls may spin and become hollow in the center after acquiring some angular momentum \cite{Volkov:2002aj, Radu:2008pp, Campanelli:2009su, Arodz:2009ye, Benci:2010zz}, which may happen when forming from a system with nonzero angular momentum or in a Q-ball collision \cite{Hou:2022jcd}. Q-balls may naturally arise in the early universe \cite{Affleck:1984fy,Enqvist:1997si,Enqvist:1998en,Enqvist:1999mv,Kasuya:1999wu,Kasuya:2000wx,Kasuya:2001hg,Multamaki:2002hv,Harigaya:2014tla, Zhou:2015yfa, Hou:2022jcd} and are a candidate for dark matter \cite{Kusenko:1997si,Enqvist:1998xd,Banerjee:2000mb,Kusenko:2001vu,Roszkowski:2006kw,Shoemaker:2009kg,Kasuya:2011ix,Kasuya:2012mh,Kawasaki:2019ywz}. In the presence of strong gravity effects, the Q-ball counterparts are known as boson stars \cite{Kaup:1968zz, Liebling:2012fv, Schunck:2003kk}, which are another candidate for dark matter \cite{Liebling:2012fv, Maselli:2017vfi}. Q-balls can also be made in laboratories \cite{Enqvist:2003zb, Bunkov:2007fe} and can have composite charge structures \cite{Battye:2000qj, Copeland:2014qra, Xie:2021glp}. 

Superradiance, coined by Dicke \cite{Dicke:1954zz} originally for emission enhancement in a coherent medium, is a collection of phenomena where radiation is amplified during a physical process; see \cite{Bekenstein:1998nt, Brito:2015oca} for a review. The well-known Cherenkov radiation is an example of inertial motion superradiance~\cite{Ginzburg:1945zz}. Later, Zel'dovich pointed out that rotating objects, such as a radiation-absorbing cylinder or a Kerr black hole, can also superradiate \cite{Zeld1, Zeld2} (also \cite{Misner:1972kx} independently for black holes). Black hole superradiance has since been extensively studied \cite{Brito:2015oca}, thanks to its relevance to gravitational theories, astrophysics and particle physics (see {\it e.g.} \cite{Andersson:1999wj, Cardoso:2004hs, Dolan:2007mj, Hartnoll:2008vx, Casals:2008pq, Arvanitaki:2009fg, Arvanitaki:2010sy, Pani:2012vp, Witek:2012tr, Herdeiro:2013pia, Yoshino:2013ofa, Brito:2014wla, Richartz:2014lda, Benone:2015bst, Wang:2015fgp, Konoplya:2016hmd, Hod:2016iri, Baryakhtar:2017ngi, East:2017ovw, Rosa:2017ury, Cardoso:2018tly, Degollado:2018ypf, Ghosh:2018gaw, Zhang:2020sjh, Mehta:2021pwf}). Also, several novel superradiance effects have recently been observed in laboratories \cite{meinardi2003superradiance, torres2017rotational, angerer2018superradiant, kim2018coherent, luo2019electrically}.

In this letter, we shall point out that Q-balls can also superradiate, a property unknown so far, despite the long history of Q-balls. Q-ball superradiance originates from the fact that a Q-ball field is complex, having two components, and that the phase of the Q-ball solution evolves in time. Indeed, a Q-ball can be viewed as a localized Bose-Einstein condensate of particles that oscillate coherently and, as we shall see, enhances scalar waves incident on it, somewhat parallel to Dicke's original scenario. If the Q-ball acquires some angular momentum and spins, additional rotational superradiance can provide further enhancement.
Interestingly, Q-ball superradiance occurs despite the fact that the particle number is conserved in the scattering. The mass gap required for a Q-ball to exist splits the superradiance spectrum into two separate parts.

We will take the U(1) symmetric theory
\begin{align}
    \mathcal{L}&=-\eta^{\mu\nu}\f{\del\bar\Phi^*}{\del \bar x^\mu} \f{\del \bar\Phi}{\del \bar x^\nu}-V(|\bar \Phi(\bar x)|)
\end{align}
with potential $V(|\bar \Phi|) = \mu^2 |\bar \Phi|^2 - \li |\bar\Phi|^4 + \bar g |\bar\Phi|^6$ as the fiducial example.  Requiring $\bar \Phi=0$ to be the true vacuum imposes the following conditions: $\li>0$ and $\mu^2 \bar g\geq \li^2/4$ \cite{Coleman:1985ki}. Upon using the dimensionless variables $x=\mu\bar x$, $\Phi=\li^{1/2}\bar\Phi/\mu$ and $g=\mu^2\bar g/\li^2$, the model can be re-written as
\begin{align}
   \label{startLag}
    \mathcal{L}&=-\del_\mu\Phi^*\del^\mu\Phi-V(|\Phi|), ~V=|\Phi|^2-|\Phi|^4+g|\Phi|^6 .
\end{align}
This may be viewed as a low energy truncation of an effective field theory expansion for a generic U(1) scalar field (and is renormalizable in 2+1D).
In the following, we will for simplicity mainly focus on the 2+1D case, except towards the end where the 3+1D case is briefly discussed.

In polar coordinates, a general Q-ball configuration takes the form
\begin{align}
\label{Phi0}
\Phi_Q(t,r,\varphi)&=\frac{1}{\sqrt{2}}f(r)e^{-i\omega_Q t+i m_Q\varphi}
\end{align}
where $\oi_Q$ is the oscillation frequency of the Q-ball in field space, $\varphi$ is the azimuth angle and the (real space) angular phase velocity of this configuration is $\Omega_Q=\omega_Q/m_Q$ if $m_Q$ is nonzero. The (2D) non-rotating Q-ball is the special case where $m_Q=0$, in which case $f(r)$ peaks at $r=0$ and falls off quickly to zero at spatial infinity. For $m_Q\neq 0$, $f(r)$ will peak at some finite $r$ and asymptotes to zero both when $r\to 0$ and $r\to \infty$. The regularity condition at the origin requires that $f (r\to 0) \propto r^{|m_Q|}$. Without loss of generality, we assume that $\oi_Q>0$ (and also $m_Q>0$ for a spinning Q-ball). For a Q-ball to exist, $\oi_Q$ must be in the range of $({1-(4g)^{-1}})^{1/2}\lesssim\oi_Q<1$ (see Supplemental Material). Q-balls in isolation are classically stable against small perturbations \cite{Coleman:1985ki, Radu:2008pp}. However, in this letter, we will show that in ``dirty" environments where waves are scattered around, energy can actually be extracted from a Q-ball via superradiant scattering.

To this end, let us look at the perturbative equations of motion around the Q-ball solution $\Phi = \Phi_Q + \phi(t,r,\varphi)$:
\begin{align}
\label{phidotdotEq}
    \Box\phi -U(r)\phi  - e^{-2i(\oi_Q t-m_Q \varphi)} W(r)\phi^{*} &=0 ,
\end{align}
where $U=\frac{1}{2}(\f{\ud^2 V}{\ud f^2} +\frac{1}{f}\f{\ud V}{\ud f})$ and $W=\frac{1}{2}(\f{\ud^2 V}{\ud f^2} -\frac{1}{f}\f{\ud V}{\ud f})$. We see that the perturbative field $\phi$ interacts with the coherent background of the Q-ball that oscillates temporally and angularly. Indeed, as we shall see later, the Q-ball condensate can enhance the energy,  angular momentum and charge of waves incident on it, giving rise to {\it superradiance}. However, $U$ and $W$ depend on $r$, preventing a straightforward spatial Fourier decomposition.

\begin{figure}
    \centering
    \includegraphics[width=3.45in]{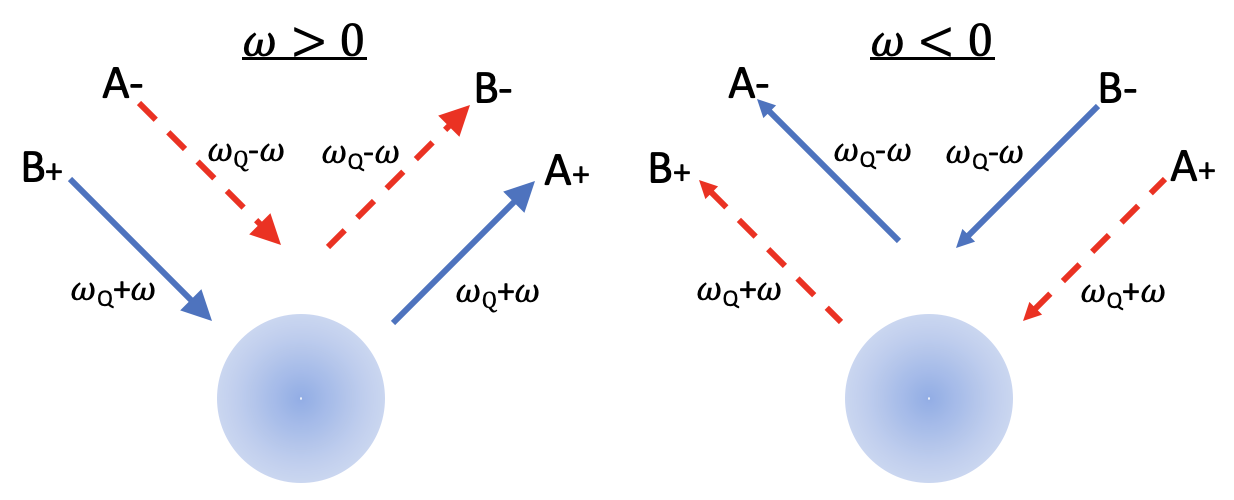}
    \caption{Ingoing and outgoing waves scattering on and off a Q-ball (cf.~Eqs.~(\ref{phidotdotEq}), (\ref{phimodes}) and (\ref{etaLargeR})). Solid (dashed) lines represent positive (negative) charge. }
    \label{fig:QballWaves}
\end{figure}

To proceed, we restrict to the minimal case where the field has only two frequencies  
\be
\label{phimodes}
\small \phi= \ei_+(\omega,m,r)e^{-i \omega_+ t+i m_+\varphi}+\ei_-(\omega,m,r)e^{-i \omega_- t+i m_-\varphi}  ,
\ee
where $\omega_\pm =  \omega_Q\pm \oi$ and $m_\pm = m_Q\pm m$. The general solution is a linear superposition of this case. The reason why two modes are needed is due to the coupling between $\phi$ and $\phi^*$, unlike the case of superradiance with a real scalar where a single frequency suffices. The mode functions satisfy the following equations
\bal
\label{etaFullEq}
\ei''_\pm + \f1r \ei'_\pm  + \(\oi_\pm^2-U -\f{m_\pm^2}{r^2} \)\ei_\pm - W\ei_\mp^* &=0 ,
\eal 
where the prime is a derivative w.r.t.\ $r$. Since $f(r\to \infty)=0$, we have $U(r\to \infty) = 1, ~W(r\to \infty) =0$. At large $r$, the above equations reduce to $\ei''_\pm + \f1r \ei'_\pm  + (\oi_\pm^2-1)\ei_\pm   \to 0$, which are solved by
\be
\label{etaLargeR}
\ei_\pm(\oi,m, r\to \infty) \to \f{A_\pm}{\sqrt{k_\pm r}} e^{i k_\pm r} + \f{B_\pm}{\sqrt{k_\pm r}} e^{-ik_\pm r}  ,
\ee
where wave numbers $k_\pm=(\oi_\pm^2-1)^{1/2}$. Here we are interested in waves scattering on and off the Q-ball. Assuming both of the two modes are propagating waves imposes the reality conditions on $k_\pm$: $|\oi_Q\pm \oi|>1$. 
The 1 on the RHS of this inequality is due to the mass gap in the scalar theory (essentially in our units the scalar mass $\mu=1$). Since we have assumed that $\oi_Q>0$, this implies that the physical boundary of $\oi$ is
\be
\label{omegaRange}
|\oi|>\oi_Q+1 .
\ee
As shown in Figure \ref{fig:QballWaves},  if $\omega>0$, the $A_-$ and $B_+$ terms represent ingoing waves and the $A_+$ and $B_-$ terms outgoing waves, where the subscript $+$ ($-$) indicates positive (negative) charge; if $\omega<0$, the wave directions flip, and $-$ ($+$) indicates positive (negative) charge.

With these set up, we solve \eref{etaFullEq} by treating it as an initial value problem in $r$. That is, we can prepare the values of $\eta_+$ and $\eta_-$ near $r=0$ and evolve \eref{etaFullEq} to a large $r$ to obtain $A_\pm$ and $B_\pm$. Regularity near the origin requires that $\ei_\pm (\oi, m,r\to 0) \to F_\pm (k_\pm r)^{|m_\pm|}$, $F_\pm$ being complex constants, which provides the ``initial'' conditions to solve the system.  Additionally, since the system is linear, if $(\eta_+,\eta_-)$ is a solution, so is $(\zeta\eta_+,\zeta^*\eta_-)$, where $\zeta$ is a complex constant. This means that we can scale one of the constants (say $F_+$) to unity, and choose the following ``initial condition'' as $r\to 0$:
\be
\label{initialCondition}
\eta_+(m, r) = (k_+r)^{|m_+|} ,~\eta_-(m,r) = F_{-} (k_-r)^{|m_-|}.
\ee    
For every $F_{-}$, the initial value problem gives a set of $A_+,A_-,B_+,B_-$. Thus, generically, we have two waves coming in and two going out. If we want to have one ingoing wave, we can use a shooting method to tune $F_{-}$ so as to make $A_-$ vanish if $\oi>0$ or to make $B_-$ vanish if $\oi<0$. (Making the ingoing $\eta_+$ mode vanish at large $r$ does not produce new results due to the symmetry of the equations.) 
The numerical schemes used to solve the equations are detailed in Supplemental Material.

To monitor how the Q-ball alters incident waves, let us define a few quantities. Firstly, we can look at the average energy in an annular region (from $r_1$ to $r_2$) far away from the origin ($r\to \infty$):
\bal
\label{energyAnn}
    E_\circledcirc &=\frac{1}{r_2-r_1}\int_{r_1}^{r_2} r\ud r\langle |\pd_t\phi|^2+|\pd_r\phi|^2+|\phi|^2 \rangle,
    \\
    &=2\frac{\omega_+^2}{k_+}\left(|A_+|^2+|B_+|^2\right)+2\frac{\omega_-^2}{k_-}\left(|A_-|^2+|B_-|^2\right) ,\nonumber
\eal
where $r_2-r_1$ includes at least a full spatial oscillation of the longest wave, $\langle\;\rangle$ is the time average over a few oscillations and we have only kept the leading order terms. We used the perturbative field to calculate the energy because the Q-ball profile falls off exponentially at large $r$. An amplification factor may be defined as the ratio of energy in the outgoing field, compared to the ingoing,
\be
\label{ampEnergy}
    \mc{A}_E=  \(  \frac{\frac{\omega_+^2}{k_+}|A_+|^2+\frac{\omega_-^2}{k_-}|B_-|^2}{\frac{\omega_+^2}{k_+}|B_+|^2+\frac{\omega_-^2}{k_-}|A_-|^2} \)^{{\rm sign}(\oi)}.
\ee
Secondly, we can also monitor how the angular momentum of the wave changes in the scattering. The angular momentum density is $T^t{}_\varphi=- \del_t \Phi^*\del_\varphi\Phi - \del_\varphi\Phi^*\del_t\Phi$. After taking the average over an annular region at large $r$ and the time average, to leading order we have
\begin{align}
    L_\circledcirc&= \frac{-1}{r_2-r_1}\int_{r_1}^{r_2}r\ud r\langle \pd_t \phi^* \pd_\varphi\phi+\pd_\varphi\phi^*{} \pd_t \phi\rangle,\\
    &=2 (|A_+|^2 \!+\! |B_+|^2)\f{\oi_+ m_+}{k_+} + 2(|A_-|^2 \!+\! |B_-|^2)\f{\oi_- m_-}{k_-} . \nonumber
    \label{LAB}
\end{align}
The amplification factor for the angular momentum is then
\be
\label{ampAngMom}
    \mc{A}_L=  \(  \frac{\frac{\omega_+m_+}{k_+}|A_+|^2+\frac{\omega_-m_-}{k_-}|B_-|^2}{\frac{\omega_+ m_+}{k_+}|B_+|^2+\frac{\omega_- m_-}{k_-}|A_-|^2} \)^{{\rm sign}(\oi)} .
\ee

\begin{figure}
	\centering
	\includegraphics[width=.4\textwidth]{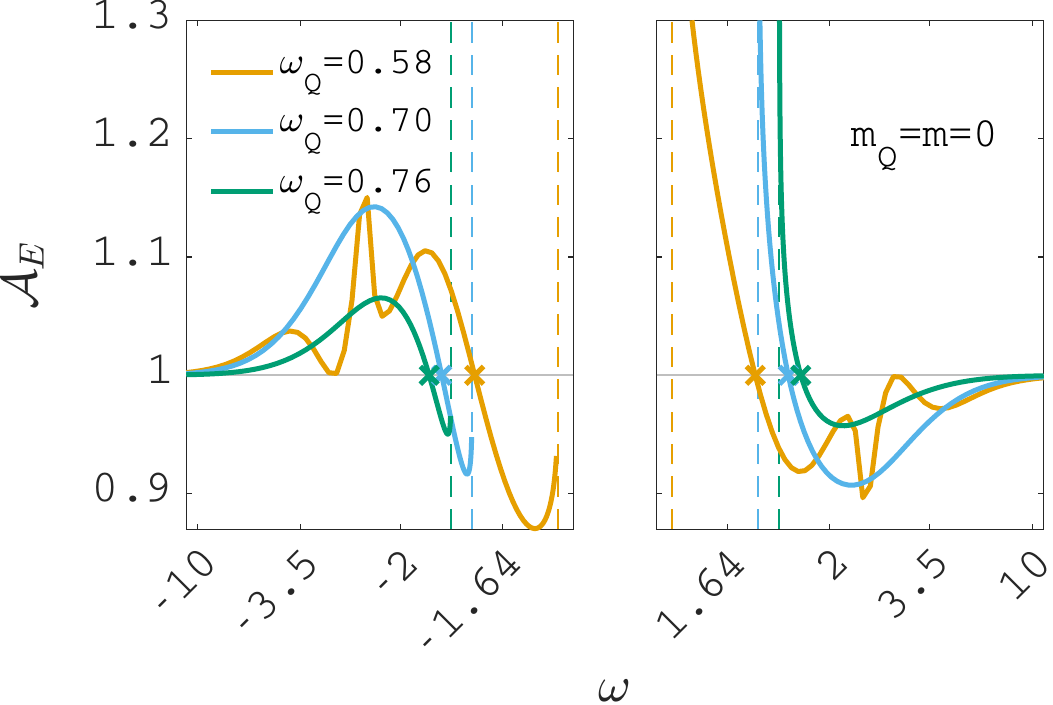}
	\caption{Amplification of the energy in the scattering for a non-spinning Q-ball with only $\eta_+(m=0)$ as the ingoing mode. The coupling is $g=1/3$. The vertical dashed lines indicate the boundary values of the reality conditions (\ref{omegaRange}), due to the mass gap. The threshold frequency $\oi_E$ for energy superradiance is given by (\ref{oiEdef}).
	}
	\label{fig:QballAmpE}
\end{figure}

One may also want to see how the particle flux changes during the scattering. 
The particle flux averaged over time and an annular region at large $r$ is given by $N_\circledcirc=2\left( |A_+|^2+|B_+|^2+|A_-|^2+|B_-|^2\right)$, and 
the amplification factor is $\mc{A}_N=\( ({|A_+|^2+|B_-|^2})/({|B_+|^2+|A_-|^2}) \)^{{\rm sign}(\oi)}$. However, due to a U(1) symmetry for the $\ei_+$ and $\ei_-$ mode in \eref{etaFullEq}, we always have $\mc{A}_N\equiv 1$. To see this, note that
the equations of motion (\ref{etaFullEq}) can be obtained from the following Lagrangian
\bal
L(\ei,\ei') &= \sum_{s=\pm} \[ |(\sqrt{r}\ei_s)'|^2 - r\(\oi_s^2\!-\!U \!-\! \f{4m_s^2-1}{4r^2}  \) |\ei_s|^2 \]
\nn
&~~~ + rW(\ei_+^*\ei_-^* + \ei_+ \ei_-)
\eal
if $r$ is viewed as ``time''. This Lagrangian is invariant under the U(1) symmetry
\be
\ei_+ = e^{i\ai} \ei_+,~~~\ei_- = e^{-i\ai} \ei_- ,~~~\ai =\mathrm{const}.
\ee
The ``Noether charge'' associated with this U(1) symmetry is 
\be
M_\ei= i r\left( \ei_{+}^*{}' \ei_{+} - \ei_{+}^*  \ei_{+}'\right) - i r\left(  \ei_{-}^*{}' \ei_{-} - \ei_{-}^*  \ei_{-}'\right) ,
\ee
which satisfies $\pd_r M_\ei=0$, meaning that $M_\ei$ is independent of $r$. At large $r$, plugging in the asymptotic solution (\ref{etaLargeR}), we get, to leading order, $M_\ei = 2(|A_+|^2 - |B_+|^2 - |A_-|^2 + |B_-|^2)$. 
On the other hand, since $\ei_\pm$ are regular at the origin $r=0$, we have $M_\ei=0$. This implies conservation of the particle number in the scattering for generic ingoing and outgoing waves: 
\be
\label{ABeqAB}
|A_+|^2+|B_-|^2= |B_+|^2+|A_-|^2 .
\ee

\begin{figure}
	\centering
	\includegraphics[width=.4\textwidth]{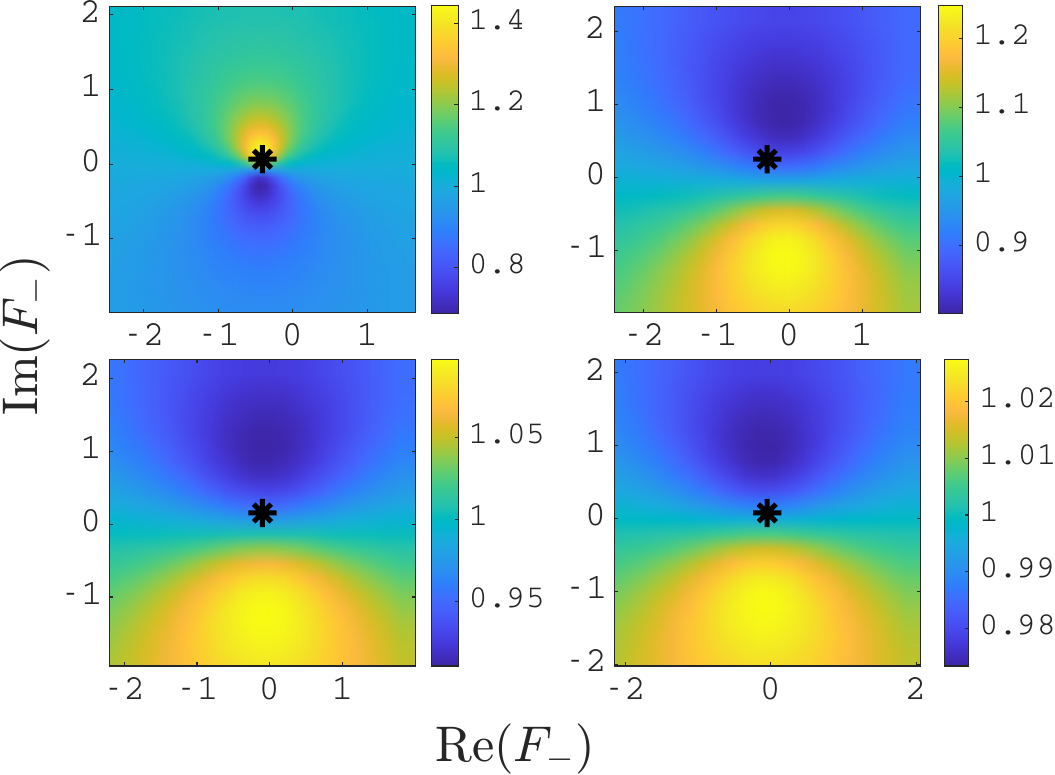}
	\caption{Amplification of the energy $\mc{A}_E$ for a non-spinning Q-ball with both $\eta_+$ and $\eta_-$ as the ingoing modes ($m_Q=m=0$). For the $F_-$ ``initial'' parameter, see \eref{initialCondition}. The other parameters are $g=1/3$ and $\omega_Q=0.7$; $\omega=1.71$(top left), $2.2$(top right), $4.2$(bottom left), $8.0$(bottom right).  Asterisks denote the case with only $\eta_+$ as the ingoing mode.}
	\label{fig:AmpEinFplane}
\end{figure}

Although the particle number is conserved in the scattering, we can still have amplification or absorption of the wave energy, depending on the frequency $\oi$ as well as the Q-ball frequency $\oi_Q$. In fact, combining the constraint (\ref{ABeqAB}) and \eref{ampEnergy}, we see that the threshold frequency $\oi_E$ that delineates the amplification and the absorption of the energy, {\it i.e.}, the case of $\mc{A}_E=1$, is given by ${\oi_+^2}/{k_+}={\oi_-^2}/{k_-}$, or explicitly,
\be
\label{oiEdef}
\f{(\oi_Q+\oi_E)^2}{\sqrt{(\oi_Q+\oi_E)^2-1}}= \f{(\oi_Q-\oi_E)^2}{\sqrt{(\oi_Q-\oi_E)^2-1}} .
\ee 
That is, this is the criterion for energy superradiance to appear at $\omega_E$. Similarly, from \eref{ampAngMom}, we see that, for an angular momentum mode $m_L$, the threshold frequency $\oi_L$ that delineates the amplification and the absorption of the angular momentum, {\it i.e.}, the case of $\mc{A}_L=1$, is given by
\be
\label{rotCri}
\f{(\oi_Q+\oi_L)(m_Q+m_L) }{\sqrt{(\oi_Q+\oi_L)^2-1}}= \f{(\oi_Q-\oi_L)(m_Q-m_L)}{\sqrt{(\oi_Q-\oi_L)^2-1}} .
\ee

\begin{figure}
	\centering
	\includegraphics[width=.4\textwidth]{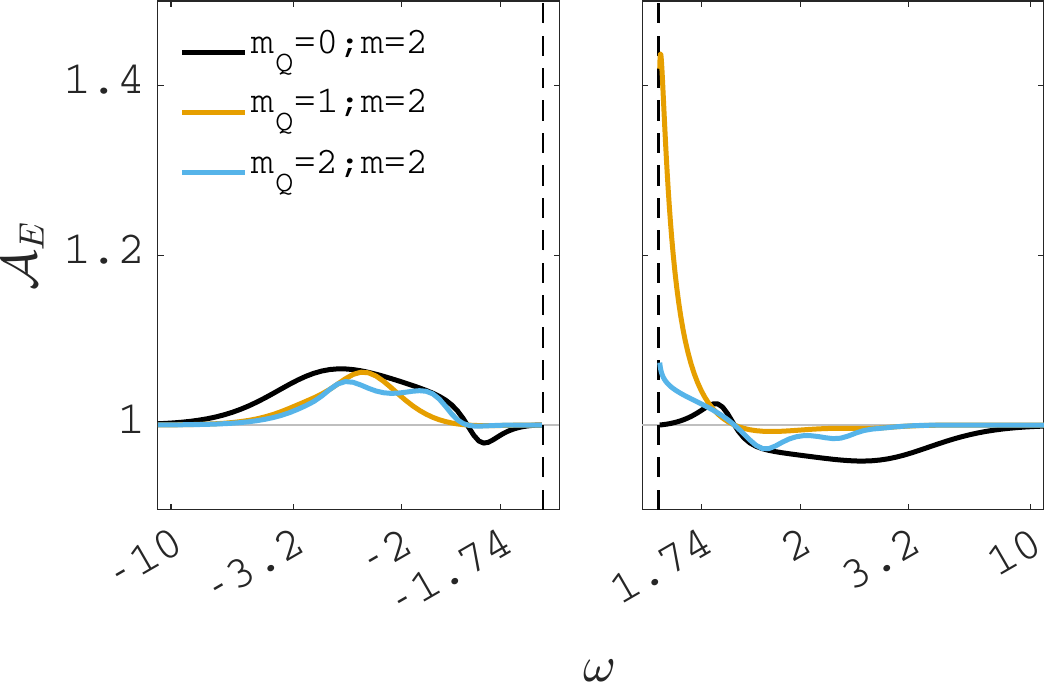}
	\caption{Amplification of the energy with only $\eta_+$  as the ingoing mode. All the Q-balls have frequency $\omega_Q=0.7$ and the coupling is $g=1/3$. The threshold frequency $\oi_E$ is still given by \eref{oiEdef}.}
	\label{fig:spinningQballAmpE}
\end{figure}

{\bf Non-spinning Q-ball} 
Let us see how the energy amplification factor $\mc{A}_E$ varies for incident frequency $\oi$ and for a few background $\oi_Q$ for the case with only an ingoing $\eta_+(m=0)$ wave; see Figure \ref{fig:QballAmpE}. (If we only have an ingoing $\eta_-$ wave, the amplification curves will just be the $\oi\to -\oi$ flip of Figure \ref{fig:QballAmpE}.) From the criterion (\ref{oiEdef}), we know that superradiance can occur when 
\be
\oi<-|\oi_{E}|~~{\rm  or} ~~\oi_Q+1<\oi<|\oi_{E}|
\ee
with $|\oi_{E}|=[1+\oi_Q^2+({1+4\oi_Q^2})^{1/2}]^{1/2}$, which is consistent with Figure \ref{fig:QballAmpE} (see Figure \ref{fig:oiEcri} in Supplemental Material for a more careful verification). 
The gaps between the positive and negative $\oi$ curves originate from the fact that the scalar has a nonzero mass, and they are different for different $\oi_Q$ (see \eref{omegaRange}). Since we send in an $\eta_+$ wave, the positive (negative) $\oi$ branch of the amplification curve is when the frequency of the ingoing wave $\oi_+=\oi_Q+\oi$ has the same (opposite) sign as the Q-ball frequency $\oi_Q$ (remember $\oi_Q>0$). We see that when the sign of $\oi_+$ is the same as $\oi_Q$, greater superradiance can be achieved, and the greatest superradiance is obtained when the incident wave has the lowest frequency or longest wavelength, that is, when the frequency $\oi$ approaches the mass gap in that branch. Typically, the peak amplification factor increases as we lower the value of $\oi_Q$, which corresponds to a bigger Q-ball. Another interesting observation is that for some $\oi_Q$ (say $\oi_Q=0.58$) there is an intriguing multi-peak structure in the amplification spectrum.

Generally, we may have both the $\eta_+$ and $\eta_-$ wave ingoing, which presumably is a typical case in a ``dirty'' environment. This mixing can be parameterized by the $F_-$ parameter. The amplifications of the energy for different mode mixings are shown in Figure \ref{fig:AmpEinFplane}. 
We see that allowing both ingoing modes in the scattering often significantly enlarges the energy amplification factor. As mentioned, this might be what one would expect, as Q-ball superradiance really arises from the interplay between the two modes of the complex scalar.

\begin{figure}
	\centering
	\includegraphics[width=.4\textwidth]{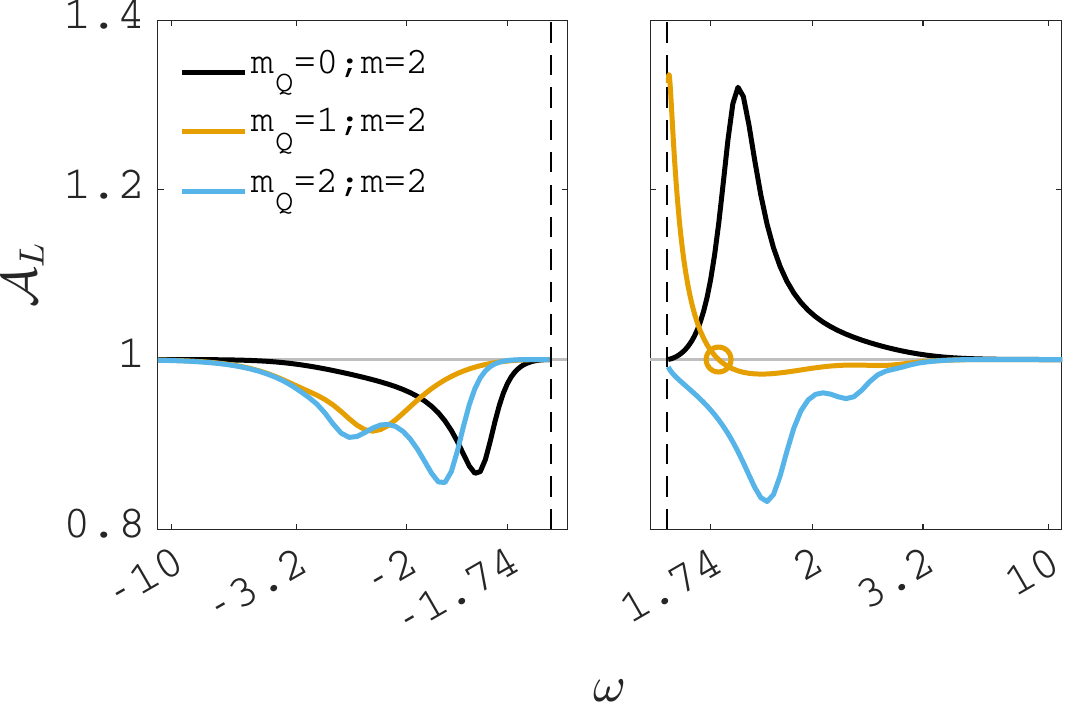}
	\caption{Amplification of the angular momentum with only $\eta_+$  as the ingoing mode. The frequencies of Q-balls are $\oi_Q=0.7$ and the coupling is $g=1/3$. The threshold frequency $\oi_L$ for angular momentum superradiance is given by \eref{rotCri}, as marked by the small circle.}
	\label{fig:spinningQballAmpL}
\end{figure}

{\bf Spinning Q-ball}  When the Q-ball spins in real space, the additional component of rotational superradiance can also be activated, which can often enhance the energy amplification when the $\eta_+$ mode rotates in the same direction as the Q-ball ({\it i.e.},  ${\rm sign}(\oi_+/m_+)={\rm sign}(\oi_Q/m_Q)$), as shown in the RHS of Figure \ref{fig:spinningQballAmpE}. On the other hand, for the opposite cases, we can see slight reductions in energy enhancement. Note that the superradiance criterion for the energy is still \eref{oiEdef}, and a non-spinning Q-ball can also induce angular momentum superradiance.

In Figure \ref{fig:spinningQballAmpL}, we plot the superradiance of the angular momentum with only the $\eta_+$ ingoing mode. We see that the amplification criterion is not $\oi<m \Omega_Q$. This is not surprising as there are two coupled modes involved in the scattering and there is Q-ball coherent superradiance on top of the rotational effects. Instead, as we have pointed out, \eref{rotCri} is the correct criterion for the angular momentum amplification. This is consistent with the case of $m_Q=1,\;m=2$ in Figure \ref{fig:spinningQballAmpL}; For the other two cases, the angular momentum is either superradiantly enhanced or reduced across the whole positive or negative $\oi$ branch, without crossing the line of $\mc{A}_L=1$, reflected in \eref{rotCri} by $\oi_L$ not having a real solution. See Supplemental Material for additional results.

\begin{figure}
	\centering
	\includegraphics[width=.4\textwidth]{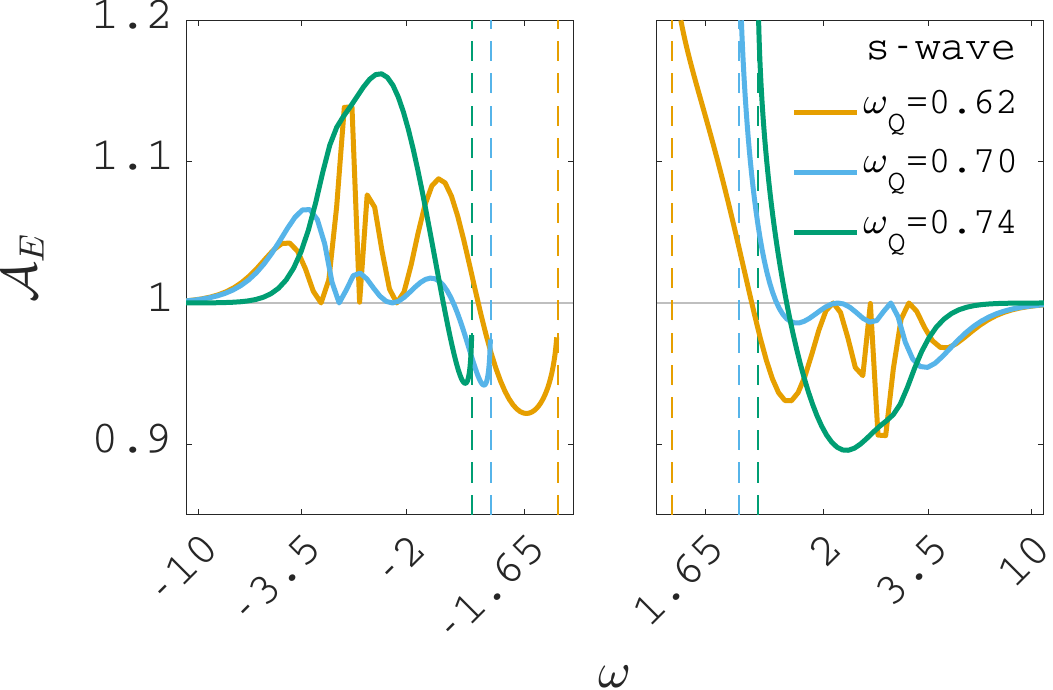}
	\caption{Amplification of the energy for a 3D spherical Q-ball with only $s$-wave $\ei_{+}$ as the ingoing mode. The coupling is $g=1/3$. We see that the 3D case tends to have more multi-peak structures.}
	\label{fig:QballAmpE3D}
\end{figure}

{\bf The 3+1D case} Finally, we shall present some first results about Q-ball superradiance in 3+1D, focusing on spherical symmetry for both the Q-ball and the scattering waves, which can be easily obtained by slightly modifying the 2+1D equations above. In Figure \ref{fig:QballAmpE3D}, we plot how the energy amplification factor varies with the frequency $\oi$. We see that the 3+1D case is completely analogous to the 2+1D case, except that the 3+1D case has more multi-peak structures in the superradiance spectra. The results of spinning Q-balls, due to its numerical complexity, will be presented elsewhere \cite{Zhang:2024ufh}.

In summary, we have found that Q-balls can induce superradiance for waves incident on them, thanks to the coherent rotation of the Q-ball in field space. For spinning Q-balls, the additional rotation in real space can further enhance superradiant emissions. An important feature in a Q-ball scattering is that it involves two coupled modes, which is essential for Q-ball superradiance to occur. Because of the presence of two coupled modes, the patterns of the superrandiance spectra are rather rich. Both the energy and angular momentum of the waves can be enhanced in the scattering, drawing energy and angular momentum from the Q-ball. However, the energy and angular momentum superradiance do not always occur simultaneously. When the ingoing waves contain both positive and negative frequency modes, the charge can also have superradiant enhancements (see Supplemental Material). Thanks to the particle number conservation in the Q-ball scattering, we have analytically identified the superradiance criteria for the energy and angular momentum. 

These results imply that superradiant amplification can take place in a ``dirty'' environment with scalar waves scattering around Q-balls. It is an interesting question whether this can be turned into spontaneous superradiant instabilities. In any case, Q-balls have long been proposed as a dark matter candidate \cite{Kusenko:1997si,Enqvist:1998xd,Banerjee:2000mb,Kusenko:2001vu,Roszkowski:2006kw,Shoemaker:2009kg,Kasuya:2011ix,Kasuya:2012mh,Kawasaki:2019ywz}, for which longevity is a prerequisite. Q-balls have also been suggested to play other interesting roles in cosmology. The existence of Q-ball superradiance begs the question of how it will affect these scenarios. 

Also, note that boson stars are essentially Q-balls in the presence of gravitational attractions. The finding in this letter hints that boson stars can also superradiate, the implications of which are worth investigating in the era of gravitational wave astronomy and other accurate gravitational observations. Indeed, Ref \cite{Gao:2023gof} has shown that the same amplification mechanism also works for boson stars. Additionally, since Q-balls have been made in laboratories, it would be interesting to observe Q-ball superradiance in condensed matter systems.

{\bf Acknowledgments} 
We would like to thank Stephen Green and Silke Weinfurtner for helpful discussions. S.Y.Z. acknowledges support from the National Natural Science Foundation of China under grant No.~12075233, No.~11947301 and No.~12047502 and from the National Key R\&D Program of China under grant No. 2022YFC220010, and is also supported by the Fundamental Research Funds for the Central Universities under grant No.~WK2030000036.  The work of P.M.S. was funded by STFC Consolidated Grant Number ST/T000732/1. 



\newpage

\section*{Supplemental material}

\subsection{More on 2+1D Q-ball superradiance}
\label{sec:2p1DMore}

\begin{figure}
	\centering
	\includegraphics[width=.35\textwidth]{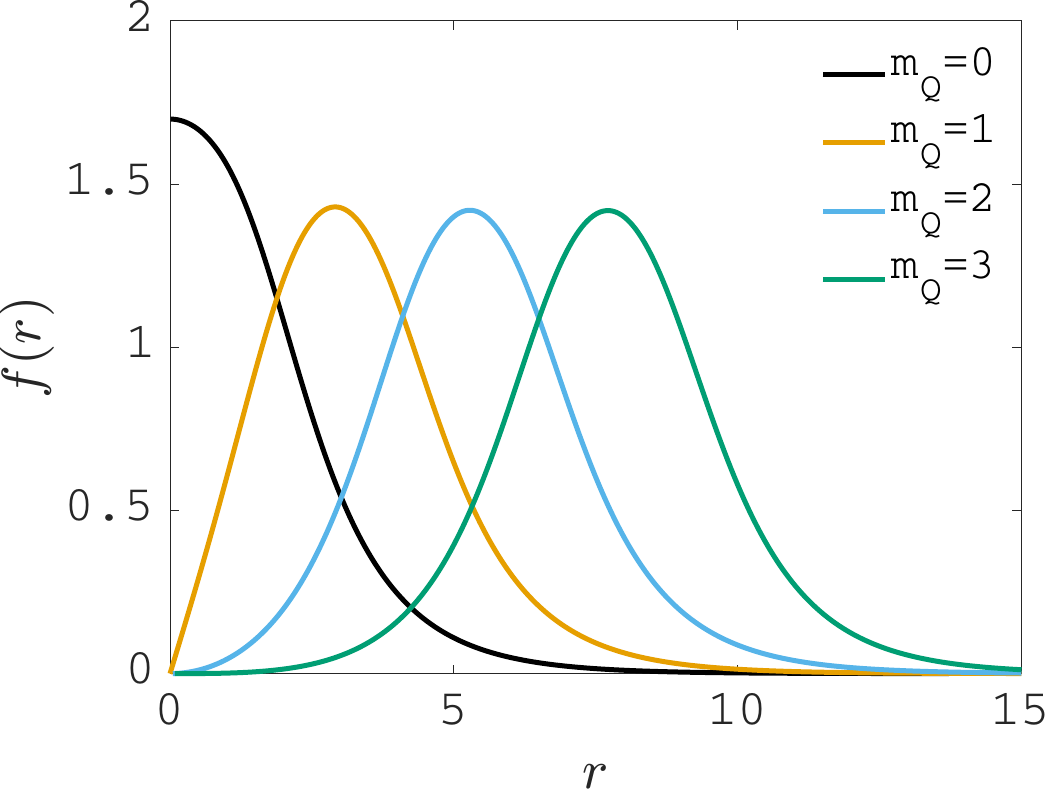}
	\caption{Q-ball profiles for the sextic potential. The parameters are $\omega_Q=0.7$ and $g=1/3$.}
	\label{fig:qball profile}
\end{figure}

Here, we shall present more results about Q-ball superradiance in 2+1D. Before that, let us first construct the radial profiles for non-spinning and spinning Q-balls. A non-spinning Q-ball in 2+1D is a disk whose density peaks in the center, while a spinning Q-ball is annulus shaped whose density peaks at some finite $r$. The Q-ball profile $f(r)$ in \eref{Phi0} can be obtained by solving the ODE  
\be
\label{eq:spinningQballEq}
\frac{\ud^2 f}{\ud r^2}+\frac{1}{r}\frac{\ud f}{\ud r}=-\frac{\del}{\del f}V_{\rm eff} ,~V_{\rm eff} \equiv \frac{\omega_Q^2}{2}f^2  -\frac{m_Q^2}{2r^2}f^2 - V ,
\ee
with the boundary conditions
\be
f(r\to0)\propto(\kappa r)^{|m_Q|} ,~~~  f(r\to\infty)\propto\frac{e^{-\kappa r}}{\sqrt{\kappa r}}, 
\ee
where $\kappa=(1-\oi_Q^2)^{1/2}$. A shooting procedure fixes the solution uniquely. A few examples of the $f(r)$ profiles are shown in Figure \ref{fig:qball profile}. 

With the above boundary conditions, for the Q-ball solution to exist, we clearly need $\oi_Q<1$. That is, $\oi_Q$ should be smaller than the scalar mass. $\oi_Q$ can not be arbitrarily small either. To find the lower limit, we can view \eref{eq:spinningQballEq} as an equation of motion for a mechanical problem with ``time'' $r$ and effective potential $V_{\rm eff}$. For large $r$, we can neglect the friction term $\frac{1}{r}\frac{\ud f}{\ud r}$ and the energy pumping term $-{m_Q^2f^2}/{2r^2}$ and have $V_{\rm eff}\simeq \f12 \omega_Q^2f^2  - V$. For the mechanical problem to have a solution, we need two local maxima in $V_{\rm eff}$ so that the ``point mass'' can: (1) roll from a finite $f$ initial to $f=0$ when $r\to \infty$ for the case of a non-spinning Q-ball; or (2) roll from $f(r\to0)=0$ initially to a finite $f$ and then roll back to $f=0$ when $r\to \infty$ for the case of a spinning Q-ball. For this to happen, we need $V_{\rm eff}$ to have a maximum at some finite $f$, which requires $\omega_Q^2$ be greater than the minimum of $2V/f^2$. This gives the condition $\oi_Q\gtrsim (1-(4g)^{-1})^{1/2}$ for our model (\ref{startLag}).

\begin{figure}
	\centering
	\includegraphics[width=.35\textwidth]{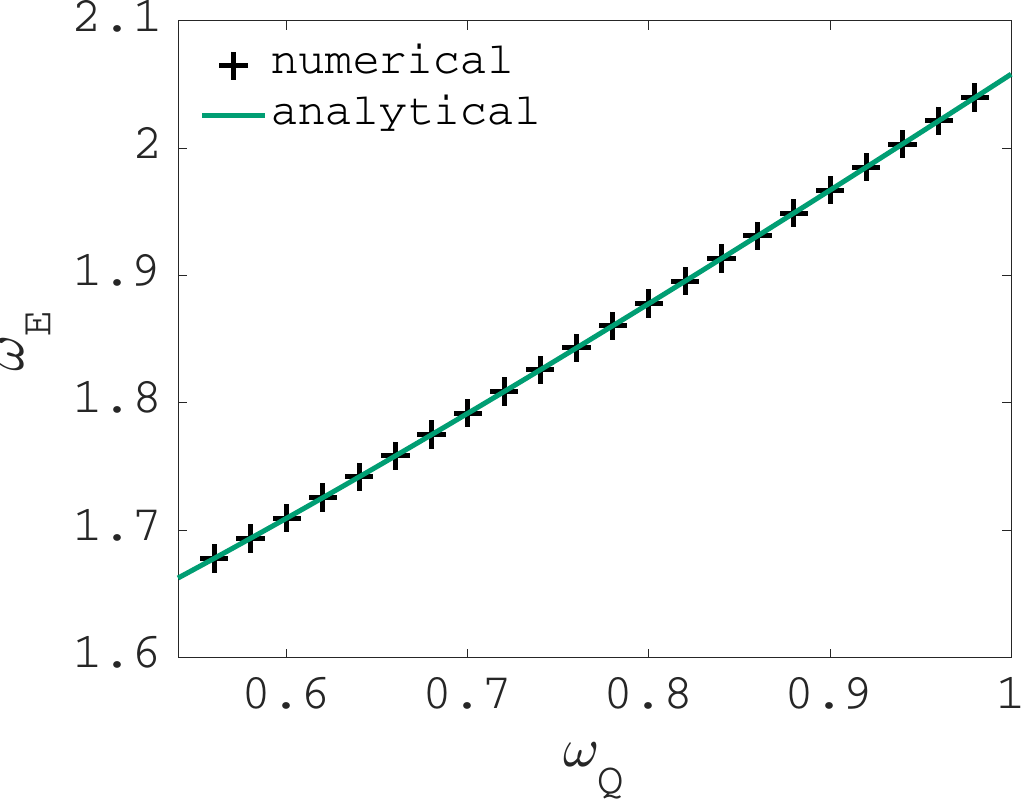}
	\caption{Numerical $\oi_E$ against the criterion (\ref{oiEdef}) $|\oi_{E}|=[1+\oi_Q^2+({1+4\oi_Q^2})^{1/2}]^{1/2}$ for a spherical Q-ball with only $\ei_+$ as the ingoing mode. The coupling is $g=1/3$.}
	\label{fig:oiEcri}
\end{figure}

As discussed in the main text, the threshold frequency $\oi_E$ for the energy to be amplified in the scattering is given by \eref{oiEdef}. To be prudent, let us verify this criterion numerically, which might also be taken as a test of the reliability of our numerical code. In Figure \ref{fig:oiEcri}, we plot $\oi_E$ obtained numerically against that of \eref{oiEdef}. Specifically, the numerical $\oi_E$ are obtained by interpolating data around $\mc{A}_E=1$, which in turn are obtained by the shooting method. We see that the agreement is rather good.

\begin{figure}
	\centering
	\includegraphics[width=.4\textwidth]{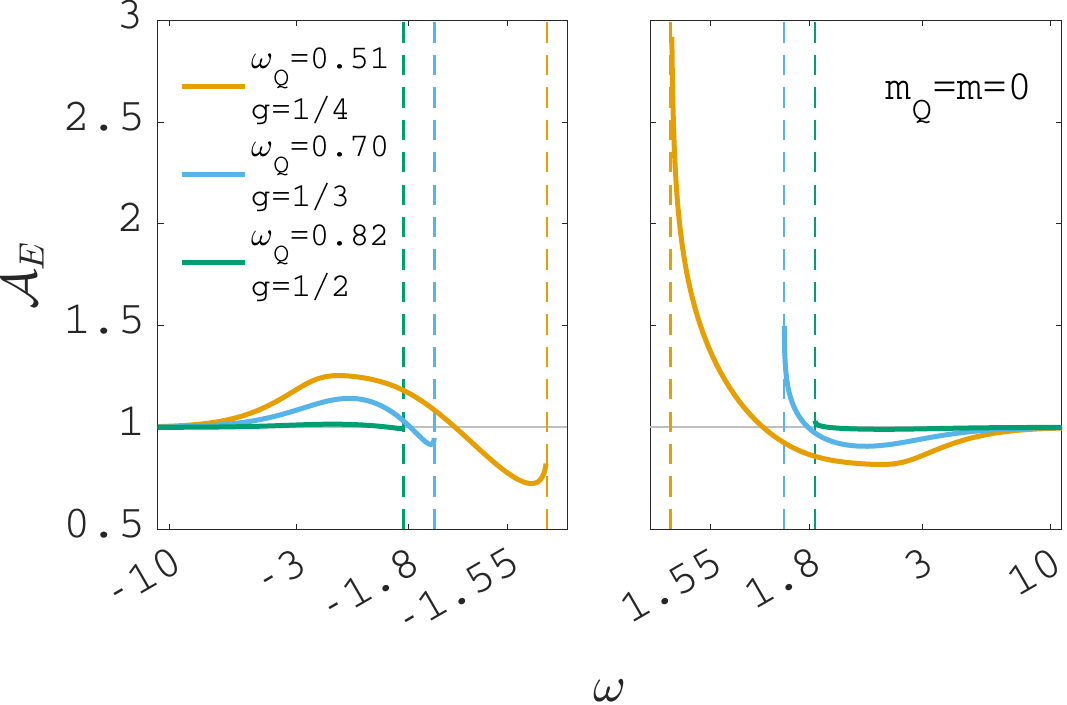}
	\caption{Amplification of the energy for non-spinning Q-balls with only $\eta_+(m=0)$ as the ingoing mode for different couplings $g$.  Note that, as done in similar figures in the main text, a nonlinear $\oi$ axis is used to showcase the fast growing parts of the curves that are close to the mass gaps. The three Q-balls have the same charge $Q=27.4$. 
	}
	\label{fig:QballAmpEg}
\end{figure}

The amplification of the energy varies significantly with the coupling constant $g$, as shown in Figure \ref{fig:QballAmpEg}. Apart from varying $g$, we also vary $\oi_Q$ so as to make sure the total charge $Q$ of the spherical Q-balls is the same.

\begin{figure}
	\centering
	\includegraphics[width=.4\textwidth]{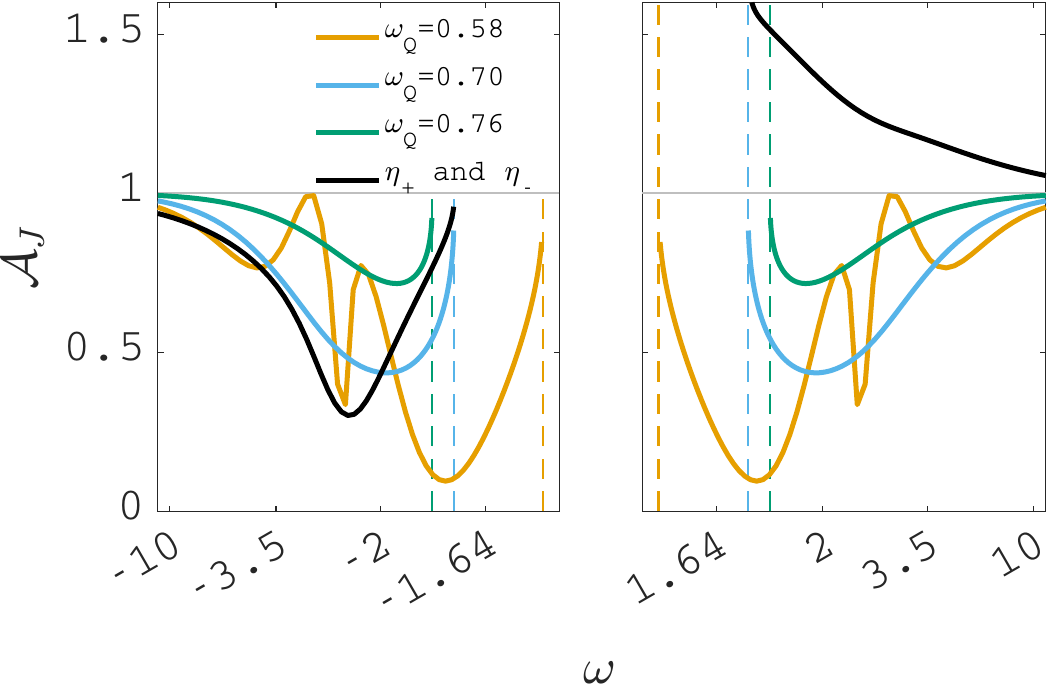}
	\caption{Amplification of the current for non-spinning Q-balls. The green, blue and orange lines are for $\ei_+(m=0)$ as the only ingoing mode. The black line is a case when both $\ei_+(m=0)$ and $\ei_-(m=0)$ are present as the ingoing modes, with $F_-=-0.5-0.1i$ and $\omega_Q=0.70$. The case with both $\ei_+$ and $\ei_-$ can have charge superradiance. The coupling is $g=1/3$. }
	\label{fig:QballAmpJ}
\end{figure}

Apart from the energy and angular momentum changes during the scattering, an additional quantity we can monitor is the radial charge flux into the Q-ball $j_r$, where $j_r$ is the radial component of the conserved current associated with the U(1) symmetry in Lagrangian (\ref{startLag}): $j_\mu=i(\Phi^*\del_\mu\Phi-\del_\mu\Phi^*\Phi)$. Again, taking the average over an annular region at large $r$ and also the time average, to leading order approximation, we get
\begin{align}
J_r^\circledcirc&= \frac{i}{r_2-r_1}\int_{r_1}^{r_2}r\ud r\langle \phi^*\pd_r\phi-\pd_r\phi^*\phi\rangle,\nn
&=2\left( -|A_+|^2+|B_+|^2-|A_-|^2+|B_-|^2\right) . 
\label{JrAB}
\end{align}
(Since both one positive charge and one negative charge give rise to one particle number, the particle flux averaged over time and an annular region at large $r$ is given by 
\be
N_\circledcirc=2\left( |A_+|^2+|B_+|^2+|A_-|^2+|B_-|^2\right),
\ee
as quoted in the main text.) For $\omega>0$, the $A_+$ and $B_+$ part correspond to positive charges (outgoing and ingoing waves respectively), while the $A_-$ and $B_-$ correspond to negative charges (ingoing and outgoing waves respectively); For $\omega<0$, then $A_+$ and $B_+$ correspond to negative charge waves (ingoing and outgoing respectively), while $A_-$ and $B_-$ correspond to positive charge waves (outgoing and ingoing respectively). So the amplification factor of the radial charge flux is then
\begin{align}
\mc{A}_J&=
\( \frac{|A_+|^2-|B_-|^2}{|B_+|^2-|A_-|^2} \)^{{\rm sign}(\oi)} .
\end{align}
Numerical results show that the $J_r^\circledcirc$ current never gets amplified for scatterings with only one ingoing mode; see Figure \ref{fig:QballAmpJ}. This can actually be inferred analytically, thanks again to the particle number conservation.  Let us consider $\omega>0$ and $A_-=0$ (the only ingoing mode being $\eta_+$) for example. The particle number conservation is then $|A_+|^2+|B_-|^2= |B_+|^2$. Combining it with the definition of $\mc{A}_J$, clearly, we must have $|\mc{A}_J|\leq1$. Similarly, the same conclusion also holds for other cases with only one ingoing mode. 
However, when both $\ei_+$ and $\ei_-$ are present as the ingoing modes, we can have superradiant enhancement for the $J_r^\circledcirc$ current; see the black line in Figure \ref{fig:QballAmpJ}.

\begin{figure}
	\centering
	\includegraphics[width=.33\textwidth]{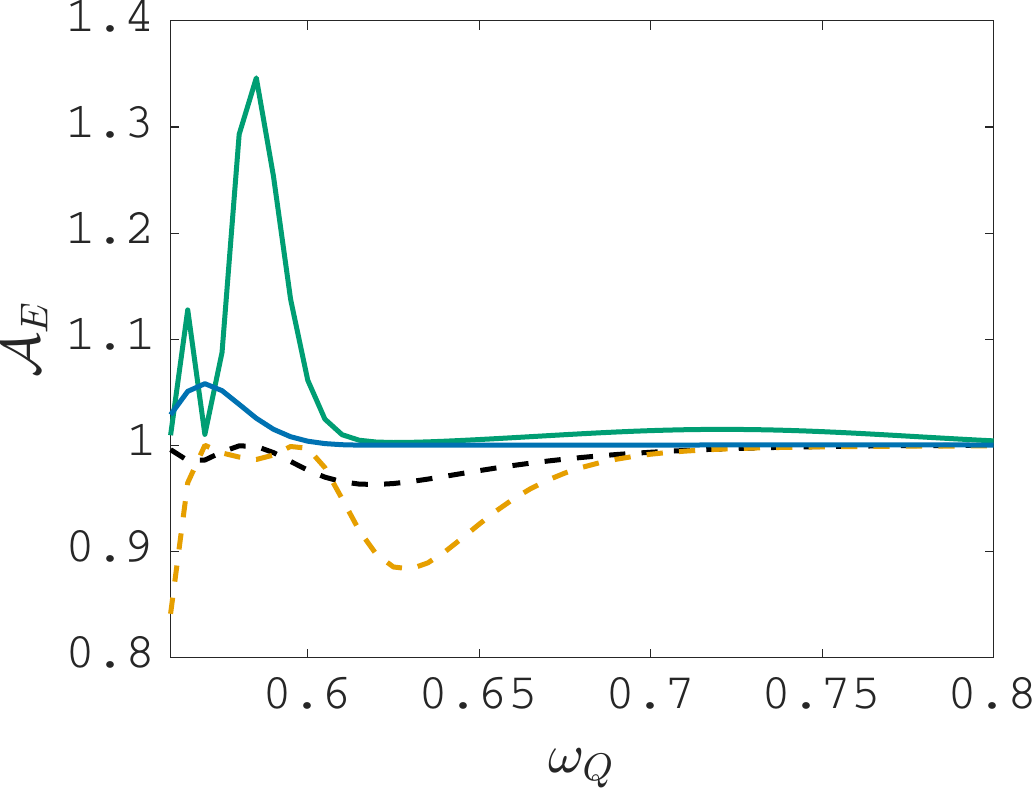}
	\includegraphics[width=.33\textwidth]{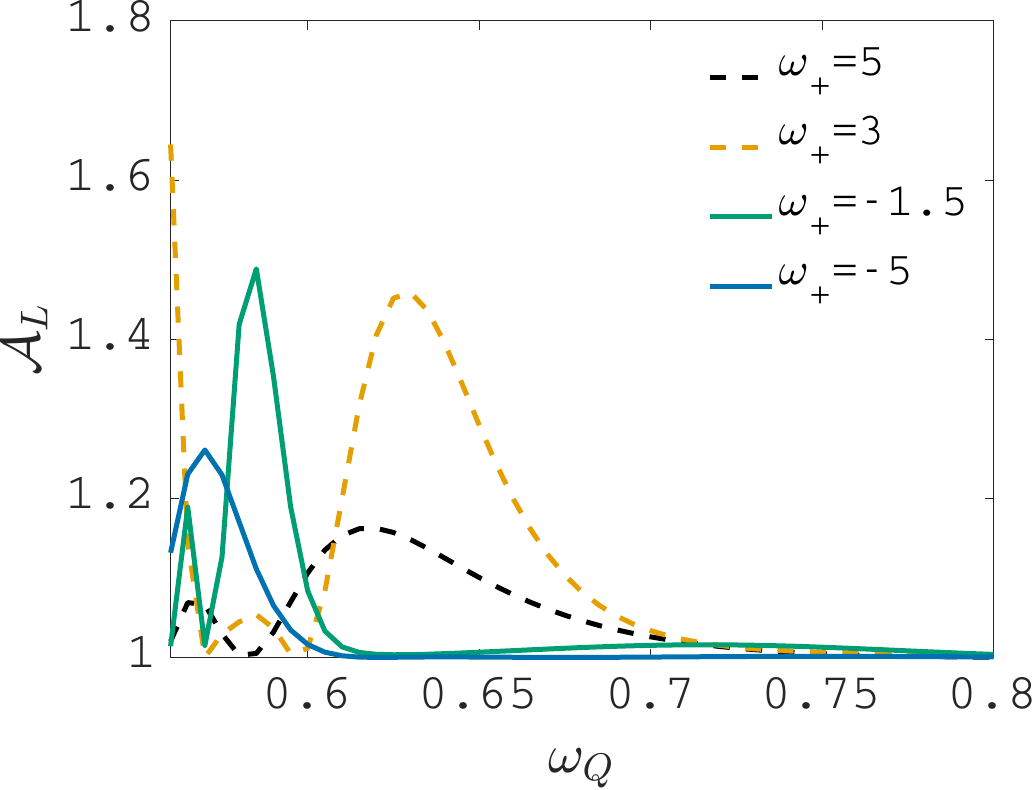}
	\caption{Amplification of the energy (top) and angular momentum (bottom) against the frequency of the Q-ball $\oi_Q$ with $m_Q=1$. $\eta_+$ is the only ingoing mode with $m_+=-2$ and various frequencies $\oi_+$. The coupling is $g=1/3$.}
	\label{fig:differentBackground}
\end{figure}

\begin{figure}
	\centering
	\includegraphics[width=.33\textwidth]{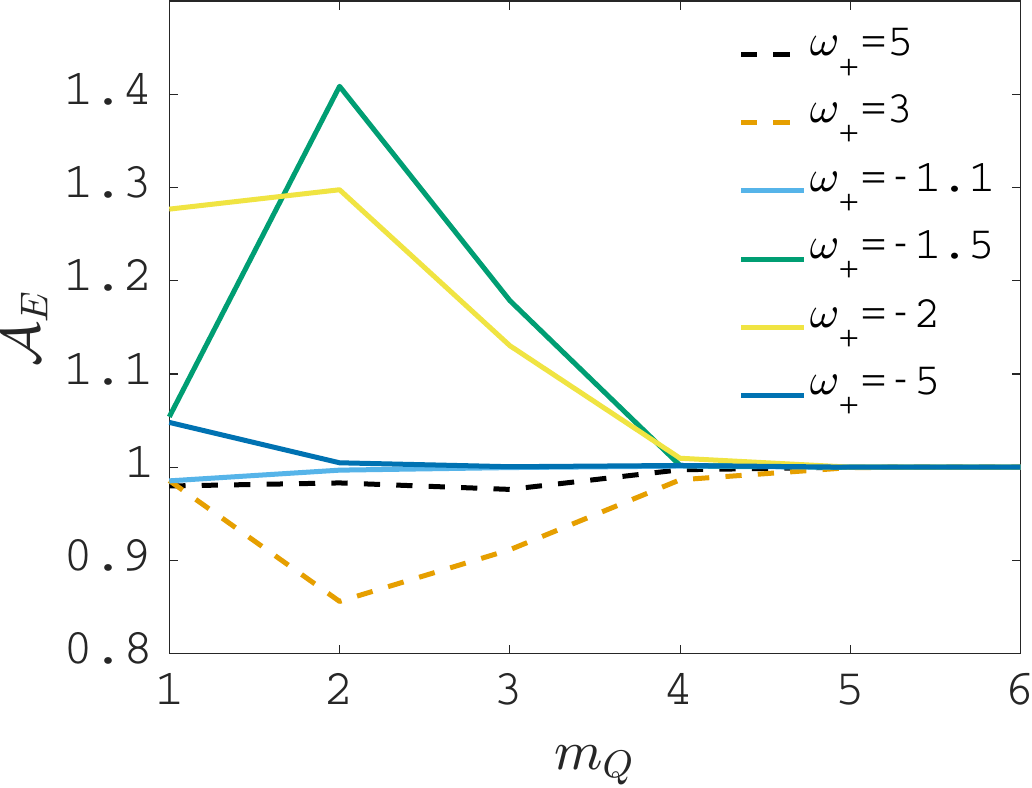}
	\includegraphics[width=.33\textwidth]{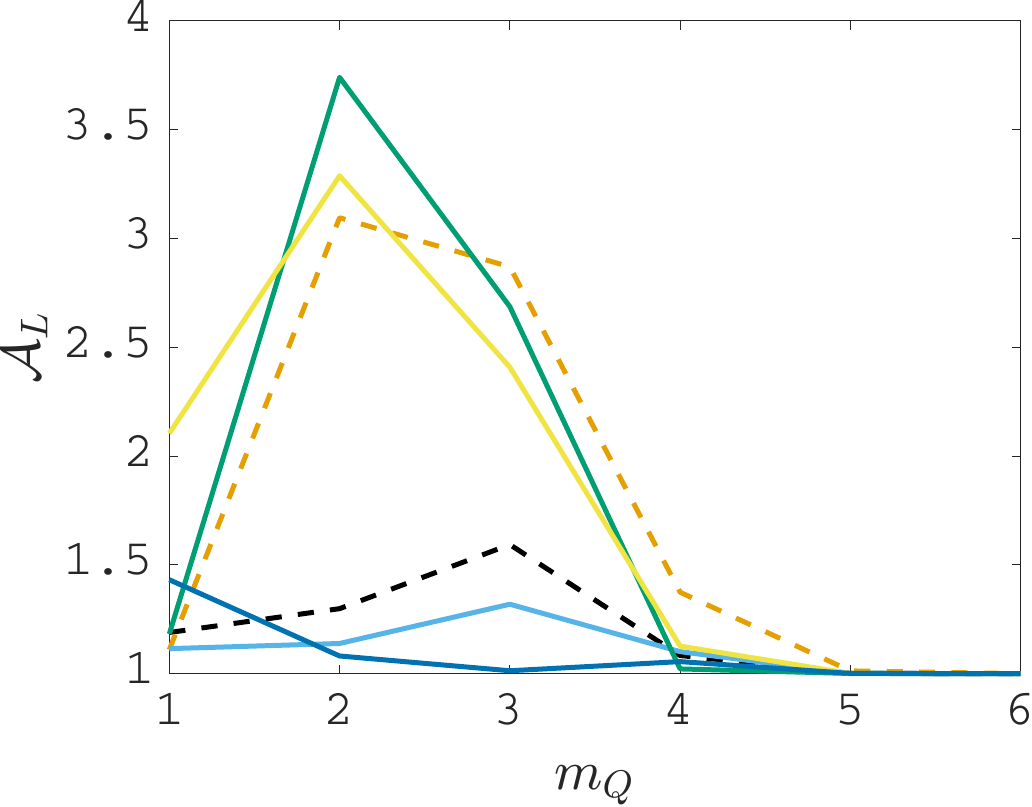}
	\caption{Amplification of the energy (top) and angular momentum (bottom) against $m_Q$ of the Q-ball with total charge $Q=270$. $\eta_+$ is the only ingoing mode with $m_+=-1$ and various frequencies $\oi_+$. The coupling is $g=1/3$.}
	\label{fig:differentBackround_Q}
\end{figure}

In Figure \ref{fig:differentBackground} and Figure \ref{fig:differentBackround_Q}, we show how the amplification factors $\mc{A}_E$ and $\mc{A}_L$ change with the background Q-ball parameters $\oi_Q$ and $m_Q$ for a single ingoing wave ($\eta_+$).
In Figure \ref{fig:differentBackground}, the angular mode numbers are fixed: $m_+=-2$ and $m_Q=1$. We see that Q-balls with smaller $\omega_Q$ tend to generate greater scattering effects, enhancement or reduction. This is not surprising since $\omega_Q$ corresponds to a bigger Q-ball. When the direction of the (real space) angular phase velocity of the $\eta_+$ mode (or ${\rm sign}(\omega_+/m_+)$) is opposite to the angular phase velocity of the Q-ball, the energy can be reduced while the angular momentum can be enhanced in the scattering. On the other hand, when the ingoing wave rotates in the same direction as that of the Q-ball, the energy and the angular momentum tend to be amplified at the same time. Also, there can be more than one peak in the $\omega_Q$ spectrum, and the impact of the Q-ball on the waves subsides for large $|\omega_+|$. 
In Figure \ref{fig:differentBackround_Q}, the spinning Q-balls share the same charge $Q=270$ but have different $m_Q$, which means that their frequencies $\omega_Q$ are also different. Notice that the angular momentum of the Q-ball is related to the total U(1) charge of the Q-ball $Q$ simply by $L=m_QQ$. We also see that there can be mismatches between the amplifications of the energy and the angular momentum. The superradiance peaks at some small $m_Q$ when the ingoing mode has a small $m_+$.

\subsection{Numerical setup}
\label{sec:NumSet}

Both the Q-ball profile equation (\ref{eq:spinningQballEq}) and the perturbative equations (\ref{etaFullEq}) are solved with {\tt Matlab}'s ordinary differential equation solver {\tt ode45}, utilizing adaptive step sizes to achieve a given error tolerance $E_T$. A shooting procedure is used to solve these equations, which is generally implemented as follows: (1) the ``initial condition'' near $r=0$ is specified by a real (complex) parameter $I$ for the profile (perturbation), using the asymptotic solution near $r=0$; (2)  {\tt ode45} is used to evolve the ODE(s) from a small $r_i$ to a large $r_f$ for a big sample of the parameter $I$, so as to evaluate the ratio between the function to be solved and its derivative, the ratio being denoted as $M$, and this gives rise to a numerical function $M_N(I)$; (For example, for perturbative modes, $M$ is $|A_-|$ if $\omega>0$ or $|B_-|$ if $\omega<0$;) (3) the ``final condition'' at $r_f$ is specified for ratio $M$ by using the asymptotic analytical solution at large $r$, which gives $M_A(r_f)$ ; (4) {\tt fzero} (for the Q-ball profiles, which have 1D parameter space) and {\tt fsolve} (for perturbative waves, which have 2D parameter space) are used to solve the algebraic equation $M_N(I)=M_A(r_f)$ to get $I=I_{sol}$, which can be used to generate the correct solution needed.

The details of the implementation of the shooting procedure differ slightly for different cases. To get the spherical Q-ball profile, the small $r$ asymptotic solution is approximated by a Taylor series up to $\mc{O}(r^6)$, with $E_T=10^{-12}$ and $r_i=0.01$ and $r_f$ determined by $f(r)$ or its derivative reaching $10^{-6}$. To get the spherical Q-ball perturbation, the small $r$ asymptotic solution is also approximated by a Taylor series up to $\mc{O}(r^6)$, with $E_T=10^{-9}$, $r_i=0.01$ and $r_f=100$. To get the spinning Q-ball profile, the small $r$ asymptotic solution is approximated by the Bessel function of the first kind, with $E_T=10^{-12}$ and $r_i$ and $r_f$ both determined by $f(r)$ or its derivative reaching $10^{-6}$. To get the spinning Q-ball perturbation, the small $r$ asymptotic solution is also approximated by the Bessel function of the first kind, with $E_T=10^{-9}$, $r_f=100$ and $r_i$ determined by $f^2|\eta_\pm|/|\pd \eta_\pm/r\pd r|$ reaching $10^{-12}$.



\begin{thebibliography}{99}

\bibitem{Rosen:1968mfz}
G.~Rosen,
J. Math. Phys. \textbf{9}, 996 (1968)

\bibitem{Friedberg:1976me}
R.~Friedberg, T.~D.~Lee and A.~Sirlin,
Phys. Rev. D \textbf{13}, 2739-2761 (1976)

\bibitem{Coleman:1985ki}
S.~R.~Coleman,
Nucl. Phys. B \textbf{262}, no.2, 263 (1985)

\bibitem{Lee:1991ax}
T.~D.~Lee and Y.~Pang,
Phys. Rept. \textbf{221}, 251-350 (1992)

\bibitem{Derrick:1964ww}
G.~H.~Derrick,
J. Math. Phys. \textbf{5}, 1252-1254 (1964)

\bibitem{Volkov:2002aj}
M.~S.~Volkov and E.~Wohnert,
Phys. Rev. D \textbf{66}, 085003 (2002)
[arXiv:hep-th/0205157 [hep-th]].

\bibitem{Campanelli:2009su}
L.~Campanelli and M.~Ruggieri,
Phys. Rev. D \textbf{80}, 036006 (2009)
[arXiv:0904.4802 [hep-th]].

\bibitem{Radu:2008pp}
E.~Radu and M.~S.~Volkov,
Phys. Rept. \textbf{468}, 101-151 (2008)
[arXiv:0804.1357 [hep-th]].

\bibitem{Arodz:2009ye}
H.~Arodz, J.~Karkowski and Z.~Swierczynski,
Phys. Rev. D \textbf{80}, 067702 (2009)
[arXiv:0907.2801 [hep-th]].

\bibitem{Benci:2010zz}
V.~Benci and D.~Fortunato,
Commun. Math. Phys. \textbf{295}, 639-668 (2010)

\bibitem{Hou:2022jcd}
S.~Y.~Hou, P.~M.~Saffin, Q.~X.~Xie and S.~Y.~Zhou,
JHEP \textbf{07}, 060 (2022)
[arXiv:2202.08392 [hep-ph]].

\bibitem{Affleck:1984fy}
I.~Affleck and M.~Dine,
Nucl. Phys. B \textbf{249}, 361-380 (1985)

\bibitem{Enqvist:1997si}
K.~Enqvist and J.~McDonald,
Phys. Lett. B \textbf{425}, 309-321 (1998)
[arXiv:hep-ph/9711514 [hep-ph]].

\bibitem{Enqvist:1998en}
K.~Enqvist and J.~McDonald,
Nucl. Phys. B \textbf{538}, 321-350 (1999)
[arXiv:hep-ph/9803380 [hep-ph]].

\bibitem{Enqvist:1999mv}
K.~Enqvist and J.~McDonald,
Nucl. Phys. B \textbf{570}, 407-422 (2000)
[erratum: Nucl. Phys. B \textbf{582}, 763-763 (2000)]
[arXiv:hep-ph/9908316 [hep-ph]].

\bibitem{Kasuya:1999wu}
S.~Kasuya and M.~Kawasaki,
Phys. Rev. D \textbf{61}, 041301 (2000)
[arXiv:hep-ph/9909509 [hep-ph]].

\bibitem{Kasuya:2000wx}
S.~Kasuya and M.~Kawasaki,
Phys. Rev. D \textbf{62}, 023512 (2000)
[arXiv:hep-ph/0002285 [hep-ph]].

\bibitem{Kasuya:2001hg}
S.~Kasuya and M.~Kawasaki,
Phys. Rev. D \textbf{64}, 123515 (2001)
[arXiv:hep-ph/0106119 [hep-ph]].

\bibitem{Multamaki:2002hv}
T.~Multamaki and I.~Vilja,
Phys. Lett. B \textbf{535}, 170-176 (2002)
[arXiv:hep-ph/0203195 [hep-ph]].

\bibitem{Harigaya:2014tla}
K.~Harigaya, A.~Kamada, M.~Kawasaki, K.~Mukaida and M.~Yamada,
Phys. Rev. D \textbf{90}, no.4, 043510 (2014)
[arXiv:1404.3138 [hep-ph]].

\bibitem{Zhou:2015yfa}
S.~Y.~Zhou,
JCAP \textbf{06}, 033 (2015)
[arXiv:1501.01217 [astro-ph.CO]].

\bibitem{Kusenko:1997si}
A.~Kusenko and M.~E.~Shaposhnikov,
Phys. Lett. B \textbf{418}, 46-54 (1998)
[arXiv:hep-ph/9709492 [hep-ph]].

\bibitem{Enqvist:1998xd}
K.~Enqvist and J.~McDonald,
Phys. Lett. B \textbf{440}, 59-65 (1998)
[arXiv:hep-ph/9807269 [hep-ph]].

\bibitem{Banerjee:2000mb}
R.~Banerjee and K.~Jedamzik,
Phys. Lett. B \textbf{484}, 278-282 (2000)
[arXiv:hep-ph/0005031 [hep-ph]].

\bibitem{Kusenko:2001vu}
A.~Kusenko and P.~J.~Steinhardt,
Phys. Rev. Lett. \textbf{87}, 141301 (2001)
[arXiv:astro-ph/0106008 [astro-ph]].

\bibitem{Roszkowski:2006kw}
L.~Roszkowski and O.~Seto,
Phys. Rev. Lett. \textbf{98}, 161304 (2007)
[arXiv:hep-ph/0608013 [hep-ph]].

\bibitem{Shoemaker:2009kg}
I.~M.~Shoemaker and A.~Kusenko,
Phys. Rev. D \textbf{80}, 075021 (2009)
[arXiv:0909.3334 [hep-ph]].

\bibitem{Kasuya:2011ix}
S.~Kasuya and M.~Kawasaki,
Phys. Rev. D \textbf{84}, 123528 (2011)
[arXiv:1107.0403 [hep-ph]].

\bibitem{Kasuya:2012mh}
S.~Kasuya, M.~Kawasaki and M.~Yamada,
Phys. Lett. B \textbf{726}, 1-7 (2013)
[arXiv:1211.4743 [hep-ph]].

\bibitem{Kawasaki:2019ywz}
M.~Kawasaki and H.~Nakatsuka,
JCAP \textbf{04}, 017 (2020)
[arXiv:1912.06993 [hep-ph]].

\bibitem{Kaup:1968zz}
D.~J.~Kaup,
Phys. Rev. \textbf{172}, 1331-1342 (1968)

\bibitem{Liebling:2012fv}
S.~L.~Liebling and C.~Palenzuela,
Living Rev. Rel. \textbf{15}, 6 (2012)
[arXiv:1202.5809 [gr-qc]].

\bibitem{Schunck:2003kk}
F.~E.~Schunck and E.~W.~Mielke,
Class. Quant. Grav. \textbf{20}, R301-R356 (2003)
[arXiv:0801.0307 [astro-ph]].

\bibitem{Maselli:2017vfi}
A.~Maselli, P.~Pnigouras, N.~G.~Nielsen, C.~Kouvaris and K.~D.~Kokkotas,
Phys. Rev. D \textbf{96}, no.2, 023005 (2017)
[arXiv:1704.07286 [astro-ph.HE]].

\bibitem{Enqvist:2003zb}
K.~Enqvist and M.~Laine,
JCAP \textbf{08}, 003 (2003)
[arXiv:cond-mat/0304355 [cond-mat]].

\bibitem{Bunkov:2007fe}
Y.~M.~Bunkov and G.~E.~Volovik,
Phys. Rev. Lett. \textbf{98}, 265302 (2007)
[arXiv:cond-mat/0703183 [cond-mat.soft]].


\bibitem{Battye:2000qj}
R.~Battye and P.~Sutcliffe,
Nucl. Phys. B \textbf{590}, 329-363 (2000)
[arXiv:hep-th/0003252 [hep-th]].

\bibitem{Copeland:2014qra}
E.~J.~Copeland, P.~M.~Saffin and S.~Y.~Zhou,
Phys. Rev. Lett. \textbf{113}, no.23, 231603 (2014)
[arXiv:1409.3232 [hep-th]].

\bibitem{Xie:2021glp}
Q.~X.~Xie, P.~M.~Saffin and S.~Y.~Zhou,
JHEP \textbf{07}, 062 (2021)
[arXiv:2101.06988 [hep-th]].

\bibitem{Dicke:1954zz}
R.~H.~Dicke,
Phys. Rev. \textbf{93}, 99-110 (1954)

\bibitem{Bekenstein:1998nt}
J.~D.~Bekenstein and M.~Schiffer,
Phys. Rev. D \textbf{58}, 064014 (1998)
[arXiv:gr-qc/9803033 [gr-qc]].

\bibitem{Brito:2015oca}
R.~Brito, V.~Cardoso and P.~Pani,
Physics,''
Lect. Notes Phys. \textbf{906}, pp.1-237 (2015)
2020,
ISBN 978-3-319-18999-4, 978-3-319-19000-6, 978-3-030-46621-3, 978-3-030-46622-0
[arXiv:1501.06570 [gr-qc]].

\bibitem{Ginzburg:1945zz}
V.~L.~Ginzburg and I.~M.~Frank,
J. Phys. (USSR) \textbf{9}, 353-362 (1945)

\bibitem{Zeld1} 
Ya.~B.~Zel'dovich, 
Zh. Eksp. Teor. Fiz. Pis'ma {\bf 14}, 270 (1971) 
[JETP Letters {\bf 14}, 180 (1971)].

\bibitem{Zeld2} 
Ya.~B.~Zel'dovich, 
Zh. Eksp. Teor. Fiz. {\bf 62}, 2076 (1971) 
[JETP  {\bf 35}, 1085 (1971)].

\bibitem{Misner:1972kx}
C.~W.~Misner,
Phys. Rev. Lett. \textbf{28}, 994-997 (1972)

\bibitem{Arvanitaki:2009fg}
A.~Arvanitaki, S.~Dimopoulos, S.~Dubovsky, N.~Kaloper and J.~March-Russell,
Phys. Rev. D \textbf{81}, 123530 (2010)
[arXiv:0905.4720 [hep-th]].

\bibitem{Arvanitaki:2010sy}
A.~Arvanitaki and S.~Dubovsky,
Phys. Rev. D \textbf{83}, 044026 (2011)
[arXiv:1004.3558 [hep-th]].

\bibitem{Pani:2012vp}
P.~Pani, V.~Cardoso, L.~Gualtieri, E.~Berti and A.~Ishibashi,
Phys. Rev. Lett. \textbf{109}, 131102 (2012)
[arXiv:1209.0465 [gr-qc]].

\bibitem{Cardoso:2004hs}
V.~Cardoso and O.~J.~C.~Dias,
Phys. Rev. D \textbf{70}, 084011 (2004)
[arXiv:hep-th/0405006 [hep-th]].

\bibitem{Hartnoll:2008vx}
S.~A.~Hartnoll, C.~P.~Herzog and G.~T.~Horowitz,
Phys. Rev. Lett. \textbf{101}, 031601 (2008)
[arXiv:0803.3295 [hep-th]].

\bibitem{Dolan:2007mj}
S.~R.~Dolan,
Phys. Rev. D \textbf{76}, 084001 (2007)
[arXiv:0705.2880 [gr-qc]].

\bibitem{Witek:2012tr}
H.~Witek, V.~Cardoso, A.~Ishibashi and U.~Sperhake,
Phys. Rev. D \textbf{87}, no.4, 043513 (2013)
[arXiv:1212.0551 [gr-qc]].

\bibitem{Baryakhtar:2017ngi}
M.~Baryakhtar, R.~Lasenby and M.~Teo,
Phys. Rev. D \textbf{96}, no.3, 035019 (2017)
[arXiv:1704.05081 [hep-ph]].

\bibitem{Herdeiro:2013pia}
C.~A.~R.~Herdeiro, J.~C.~Degollado and H.~F.~R\'unarsson,
Phys. Rev. D \textbf{88}, 063003 (2013)
[arXiv:1305.5513 [gr-qc]].

\bibitem{Degollado:2018ypf}
J.~C.~Degollado, C.~A.~R.~Herdeiro and E.~Radu,
Phys. Lett. B \textbf{781}, 651-655 (2018)
[arXiv:1802.07266 [gr-qc]].

\bibitem{Cardoso:2018tly}
V.~Cardoso, \'O.~J.~C.~Dias, G.~S.~Hartnett, M.~Middleton, P.~Pani and J.~E.~Santos,
JCAP \textbf{03}, 043 (2018)
[arXiv:1801.01420 [gr-qc]].

\bibitem{Brito:2014wla}
R.~Brito, V.~Cardoso and P.~Pani,
Class. Quant. Grav. \textbf{32}, no.13, 134001 (2015)
[arXiv:1411.0686 [gr-qc]].

\bibitem{Ghosh:2018gaw}
S.~Ghosh, E.~Berti, R.~Brito and M.~Richartz,
Phys. Rev. D \textbf{99}, no.10, 104030 (2019)
[arXiv:1812.01620 [gr-qc]].

\bibitem{Hod:2016iri}
S.~Hod,
Phys. Lett. B \textbf{758}, 181-185 (2016)
[arXiv:1606.02306 [gr-qc]].

\bibitem{Casals:2008pq}
M.~Casals, S.~R.~Dolan, P.~Kanti and E.~Winstanley,
JHEP \textbf{06}, 071 (2008)
[arXiv:0801.4910 [hep-th]].

\bibitem{Benone:2015bst}
C.~L.~Benone and L.~C.~B.~Crispino,
Phys. Rev. D \textbf{93}, no.2, 024028 (2016)
[arXiv:1511.02634 [gr-qc]].


\bibitem{East:2017ovw}
W.~E.~East and F.~Pretorius,
Phys. Rev. Lett. \textbf{119}, no.4, 041101 (2017)
[arXiv:1704.04791 [gr-qc]].

\bibitem{Konoplya:2016hmd}
R.~A.~Konoplya and A.~Zhidenko,
JCAP \textbf{12}, 043 (2016)
[arXiv:1606.00517 [gr-qc]].

\bibitem{Rosa:2017ury}
J.~G.~Rosa and T.~W.~Kephart,
Phys. Rev. Lett. \textbf{120}, no.23, 231102 (2018)
[arXiv:1709.06581 [gr-qc]].

\bibitem{Richartz:2014lda}
M.~Richartz, A.~Prain, S.~Liberati and S.~Weinfurtner,
Phys. Rev. D \textbf{91}, no.12, 124018 (2015)
[arXiv:1411.1662 [gr-qc]].


\bibitem{Wang:2015fgp}
M.~Wang and C.~Herdeiro,
Phys. Rev. D \textbf{93}, no.6, 064066 (2016)
[arXiv:1512.02262 [gr-qc]].

\bibitem{Yoshino:2013ofa}
H.~Yoshino and H.~Kodama,
PTEP \textbf{2014}, 043E02 (2014)
[arXiv:1312.2326 [gr-qc]].

\bibitem{Zhang:2020sjh}
C.~Y.~Zhang, S.~J.~Zhang, P.~C.~Li and M.~Guo,
JHEP \textbf{08}, 105 (2020)
[arXiv:2004.03141 [gr-qc]].

\bibitem{Andersson:1999wj}
N.~Andersson and K.~Glampedakis,
Phys. Rev. Lett. \textbf{84}, 4537-4540 (2000)
[arXiv:gr-qc/9909050 [gr-qc]].

\bibitem{Mehta:2021pwf}
V.~M.~Mehta, M.~Demirtas, C.~Long, D.~J.~E.~Marsh, L.~McAllister and M.~J.~Stott,
JCAP \textbf{07}, 033 (2021)
[arXiv:2103.06812 [hep-th]].




\bibitem{meinardi2003superradiance}
  F.~Meinardi, M.~Cerminara, A.~Sassella, R.~Bonifacio and R.~Tubino,
  Phys. Rev. Lett. \textbf{91}, no.24, 247401 (2003)

\bibitem{torres2017rotational}
  T.~Torres, S.~Patrick, A.~Coutant, M.~Richartz, E.~W.~Tedford and S.~Weinfurtner,
  Nature Phys. \textbf{13}, 833-836 (2017)

\bibitem{angerer2018superradiant}
  A.~Angerer, K.~Streltsov, T.~Astner, S.~Putz, H.~Sumiya, S.~Onoda, J.~Isoya, W.~J.~Munro, K.~Nemoto, J.~Schmiedmayer and J.~Majer,
  Nature Phys. \textbf{14}, 1168-1172 (2018)

\bibitem{kim2018coherent}
  J.~Kim, D.~Yang, S.~Oh and K.~An,
  Science \textbf{359}, no.6376, 662-666 (2018)

\bibitem{luo2019electrically}
  Y.~Luo, G.~Chen, Y.~Zhang, L.~Zhang, Y.~Yu, F.~Kong, X.~Tian, Y.~Zhang, C.~Shan, Y.~Luo, J.~Yang, V.~Sandoghdar, Z.~Dong, and J.~G.~Hou,
  Phys. Rev. Lett. \textbf{122}, no.23, 233901 (2019)

\bibitem{Zhang:2024ufh}
G.~D.~Zhang, F.~M.~Chang, P.~M.~Saffin, Q.~X.~Xie and S.~Y.~Zhou,
[arXiv:2402.03193 [hep-th]].

\bibitem{Gao:2023gof}
H.~Y.~Gao, P.~M.~Saffin, Y.~J.~Wang, Q.~X.~Xie and S.~Y.~Zhou,
[arXiv:2306.01868 [gr-qc]].


\end{thebibliography}
\end{document}